\documentclass[conference]{IEEEtran}
\usepackage[T1]{fontenc}
\usepackage{amsmath,amssymb}
\usepackage{graphicx}
\usepackage{booktabs}
\usepackage{xcolor}
\usepackage{tikz}
\usetikzlibrary{arrows.meta,positioning,fit,backgrounds,calc}
\usepackage[hidelinks]{hyperref}
\usepackage{url}
\usepackage{cite}

\newcommand{\avgFrames}{4}
\newcommand{\bandHi}{33}
\newcommand{\bandLoAbs}{31}
\newcommand{\bcnnMhz}{66.67}
\newcommand{\bcnnWns}{2.616}
\newcommand{\binaryMacsM}{37.1}
\newcommand{\boardAllExact}{639}
\newcommand{\boardAllN}{639}

\newcommand{\boardInferRate}{27}

\newcommand{\boardTOneAcc}{79.39}

\newcommand{\boardTOneN}{524}
\newcommand{\boardTTwoAcc}{79.13}

\newcommand{\boardTTwoN}{115}

\newcommand{\busyPct}{90.8}
\newcommand{\camAvgRaw}{70.7}
\newcommand{\camAvgStretch}{76.9}
\newcommand{\camFpsModel}{26.68}
\newcommand{\camlikeTOneRaw}{47.7}
\newcommand{\camlikeTOneSt}{78.2}
\newcommand{\camlikeTTwoRaw}{67.0}
\newcommand{\camlikeTTwoSt}{78.3}

\newcommand{\commitToDecisionMs}{34.2}
\newcommand{\confTOneAcc}{90.2}
\newcommand{\confTOneCov}{50.8}
\newcommand{\confTTwoAcc}{95.7}
\newcommand{\confTTwoCov}{40.0}
\newcommand{\coreBram}{10}
\newcommand{\coreEnergyMj}{0.75}

\newcommand{\coreLut}{744}
\newcommand{\corePowMw}{22}
\newcommand{\cpuBcnnOneMs}{8.13}

\newcommand{\cpuResnetOneMs}{56.39}
\newcommand{\cpuResnetSixteenMs}{42.49}

\newcommand{\cyclesInf}{2{,}269{,}477}
\newcommand{\dataCrossings}{7{,}552}
\newcommand{\dataImages}{11{,}281}
\newcommand{\detectLut}{1{,}814}
\newcommand{\devBram}{50}

\newcommand{\frameBytes}{614{,}400}
\newcommand{\frameMsModel}{37.49}
\newcommand{\frameReduction}{38{,}400}

\newcommand{\hostSharePct}{21.7}
\newcommand{\hostSixteenSharePct}{27.9}
\newcommand{\inferMs}{34.04}
\newcommand{\infoBytes}{16}

\newcommand{\mmcmPowMw}{109}

\newcommand{\powDynW}{0.221}
\newcommand{\powStaticW}{0.073}
\newcommand{\powTotalW}{0.294}
\newcommand{\precisionTOne}{93.8}
\newcommand{\precisionTTwo}{81.2}
\newcommand{\recallTOne}{67.5}
\newcommand{\recallTTwo}{83.6}

\newcommand{\resnetTTwo}{95.7}

\newcommand{\selPredTOneForty}{93.8}

\newcommand{\selPredTTwoThirty}{97.1}
\newcommand{\slackMs}{3.44}
\newcommand{\snapshotBytes}{4{,}826}
\newcommand{\snapshotReduction}{127}
\newcommand{\splitTest}{2{,}164}
\newcommand{\splitTrain}{7{,}097}
\newcommand{\splitVal}{1{,}905}

\newcommand{\sysBram}{19}
\newcommand{\sysDsp}{1}

\newcommand{\sysLut}{3{,}370}
\newcommand{\sysLutPct}{16.2}

\newcommand{\trainSources}{14}

\newcommand{\vThreeTTwo}{79.1}
\newcommand{\vThreeThr}{477}

\newcommand{\weightBits}{93{,}408}
\newcommand{\weightKB}{11.4}

\begin{document}

\title{Hardwired Distributed Processing Units for Inference:\\Neural Decisions at the Sensor that Free the Host}

\author{\IEEEauthorblockN{Raj Mali}
\IEEEauthorblockA{\textit{Independent Researcher}\\
{raj.mali.research@gmail.com}\\
ORCID: \href{https://orcid.org/0009-0001-6351-0062}{0009-0001-6351-0062}}}

\maketitle

\begin{abstract}
A camera normally sends every pixel to a host processor, which has to wait for a whole frame before it can decide anything. Hardwired distributed processing units (HDPUs) move routine work to the sensor, but so far we have used them only for pre-processing: edges, corners and a steering signal. Here we show that an HDPU can take a neural decision as well, and so lift the inference off the host. We build one for zebra-crossing detection on a low-cost Artix-7 FPGA. An OV7670 camera feeds a hardware pipeline that reduces each frame to $80\times60$ pixels, restores its contrast and evaluates a four-layer binarized CNN, with no processor, no operating system and no full-resolution frame store. The unit decides on every camera frame, in exactly \cyclesInf{} clock cycles (\inferMs~ms) whatever the image, and on the board it returns the same scores and decisions as the bit-exact software model on all \boardAllN{} test images. The whole detection path takes \detectLut{} LUTs. On street photographs, \confTTwoAcc\,\% of its answers are correct when it answers instead of reporting ``not sure'', and a hardware contrast stage lifts accuracy on camera-like dashcam frames from \camlikeTOneRaw\,\% to \camlikeTOneSt\,\%. The host runs none of this work. It is spared \cpuBcnnOneMs~ms of a core in every \frameMsModel-ms frame period, and a full-precision ResNet-18 would take longer than a frame period even on 16 threads. It also receives \infoBytes{} bytes per frame instead of \frameBytes, \frameReduction{} times less.
\end{abstract}


\section{Introduction}
Most vision systems still follow one routine. The camera sends every pixel to a host processor, the host stores the frame, and only then does a program run over it. Processors do this well, but the routine has two costs that grow with the work we ask of it. Nothing can be decided before a whole frame has arrived, and every frame takes a share of the host's time, whether or not anything interesting is in it.

In earlier work we proposed taking the routine, always-on part of that job away from the host. A \emph{hardwired distributed processing unit} (HDPU) is a small fixed datapath attached to the sensor and dedicated to one task. It works on the pixel stream as the pixels arrive, shares no bus or scheduler with its neighbours, and passes only a compact result to the host~\cite{papertwo}. Our first units, built from one library of operators that share a timing contract~\cite{paperone}, did pre-processing: Canny edges, Harris corners and a line tracker that emits a steering code. The decision itself was still left to the host.

In modern systems that decision is made by a neural network, and a network is normally run as a program. A processor, or an accelerator that the processor controls~\cite{jouppi2017}, fetches instructions and weights from memory for every input. This is flexible, but it is also where most of the energy goes~\cite{horowitz2014,hameed2010}, and the host stays busy for as long as the network runs. Industry has started to move the other way. Taalas builds chips in which the weights and dataflow of one model are part of the silicon, and reports up to 17{,}000 generated tokens per second for Llama~3.1~8B~\cite{taalas2026}; in August 2026 AMD announced that it would acquire the company~\cite{amd2026}.

This paper asks whether the sensor can take the same step. Can an HDPU not only prepare the data but also make the decision? We answer with a working unit. It detects zebra crossings, a task that matters for navigation aids for visually impaired pedestrians and for driver assistance~\cite{mascetti2016,yu2019,zhang2022}, on a low-cost Artix-7 FPGA, with a binarized CNN~\cite{courbariaux2016,rastegari2016} that replaces multiply-accumulate with XNOR and population count. Our contributions are:
\begin{enumerate}
\item \textbf{An inference HDPU.} A unit that turns camera frames into decisions at the sensor, with no processor in the loop. It decides on every frame in a fixed \cyclesInf{} clock cycles, and on the board its scores and decisions equal those of its bit-exact model on all \boardAllN{} test images (Sections~\ref{sec:unit} and~\ref{sec:board}).
\item \textbf{The sensor side of the decision.} Pre-processing, a hardware contrast stage that recovers the accuracy a real sensor costs (\camlikeTOneRaw\,\% to \camlikeTOneSt\,\% on camera-like frames), and a decision with a not-sure band that makes \confTTwoAcc\,\% of its street-photo answers correct, all in \detectLut{} LUTs (Sections~\ref{sec:unit} and~\ref{sec:acc}).
\item \textbf{A complete verification chain,} from a Python integer model through RTL simulation to the physical board, with every level compared against the same model (Section~\ref{sec:verif}).
\item \textbf{A measured account of what the host is spared:} the inference time, the share of every frame period, and the data per frame, which falls from \frameBytes{} to \infoBytes{} bytes (Section~\ref{sec:host}).
\end{enumerate}

\section{Background and Related Work}
\textbf{HDPUs.} An HDPU is a fixed datapath for one task, with no instruction stream, operating system or frame buffer, placed at the sensor in the spirit of near-sensor computing~\cite{zhou2020}. Several HDPUs can watch the same pixel stream without a common bus, arbiter or scheduler, so adding a unit costs logic but no time. A Canny unit, a Harris corner unit and a line tracker already run side by side on one camera stream and return part of a host core to the host in every frame~\cite{papertwo}, all built from $3\times3$ operators that answer exactly two clock cycles after their window~\cite{paperone}.

\textbf{Binarized networks.} A binarized network constrains weights and activations to $\pm1$, so a multiply-accumulate over $n$ inputs becomes an XNOR of two $n$-bit words followed by a population count~\cite{courbariaux2016,rastegari2016}. Batch normalisation followed by the sign function folds into one integer threshold per neuron~\cite{umuroglu2017}.

\textbf{Inference on FPGAs.} FINN builds streaming dataflow accelerators for binarized networks whose folding factors trade throughput for area~\cite{umuroglu2017,blott2018}; its CIFAR-10 prototype classifies 21{,}906 images per second at 80.1\,\% accuracy on a Zynq board~\cite{umuroglu2017}. hls4ml unrolls small networks with their weights as constants to reach sub-microsecond decisions in particle-physics triggers~\cite{duarte2018}, and LogicNets and LUTNet map neurons directly onto LUT truth tables~\cite{umuroglu2020,wang2019}. We use the same techniques. What we add is the placement: the unit sits at the camera, decides on every frame, and needs nothing around it.

\textbf{Model-specific silicon.} The TPU executes inference in a systolic array that receives its instructions and data from a host over PCIe~\cite{jouppi2017}. Groq's tensor streaming processor schedules all data movement at compile time, so its execution is deterministic~\cite{abts2020}. Taalas goes furthest and makes the weights part of the chip~\cite{taalas2026}. An inference HDPU shares the last two properties, fixed timing and weights beside the datapath, and sits at the sensor.

\textbf{Offload and baselines.} Offload pays when invoking the accelerator and moving data to it cost little relative to the work~\cite{altaf2017}, and reported speed-ups often shrink against optimised processor code~\cite{lee2010}. An HDPU is never invoked and never receives data from the host, so the cost it removes is the host's own: time and traffic.

\textbf{Zebra-crossing detection.} Published detectors run on phones~\cite{mascetti2016,yu2019} or on embedded GPU modules; CDNet reports an F1 score above 94\,\% at 33.1 frames per second on a Jetson Nano~\cite{zhang2022}. The engine of the unit described here was first designed and verified in simulation in~\cite{malireport}.

\section{An Inference HDPU at the Sensor}\label{sec:unit}

\begin{figure*}[t]
\centering
\includegraphics[width=\textwidth]{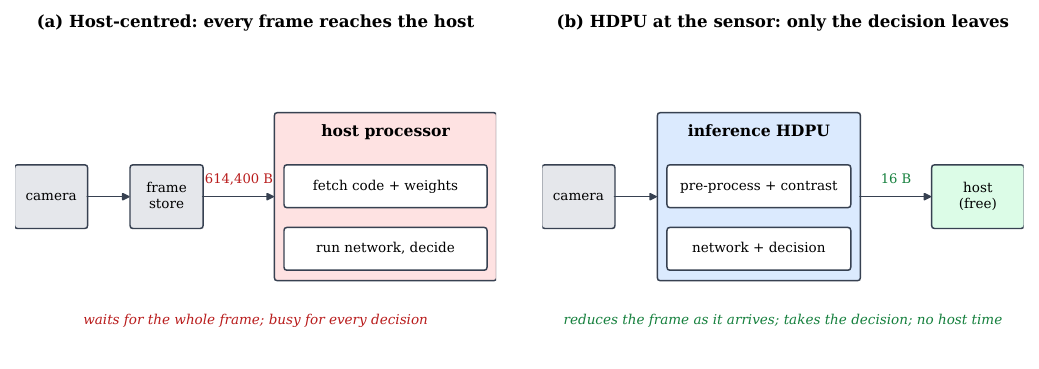}
\caption{Two ways to get a decision from a camera. (a) The host-centred routine moves every frame to the host, which fetches instructions and weights and runs the network. (b) The inference HDPU reduces the frame as it arrives and takes the decision at the sensor; the host receives \infoBytes{} bytes and does no work.}
\label{fig:org}
\end{figure*}

\subsection{Organisation}
Fig.~\ref{fig:org} contrasts the two organisations. In the host-centred one, the frame is stored, moved to the host and processed by a program, so the decision cannot exist before the frame has arrived and the host has spent its time. In the inference HDPU the unit reads the camera pins directly, in parallel with the path that displays the picture, so it never asks anyone for data. It has no instruction stream: the network's weights and thresholds sit in on-chip memory beside a fixed datapath. Its schedule is fixed, because a decision takes the same number of clock cycles for every frame, a number set by the network's layer table and by nothing else. And its output is a decision of a few bytes, not a frame.

\subsection{From pixels to the network's input}
\begin{figure*}[t]
\centering
\begin{tikzpicture}[font=\footnotesize, >=Stealth, node distance=4mm,
  b/.style={draw, rounded corners=2pt, minimum height=7mm, align=center, inner sep=2pt},
  unit/.style={b, fill=blue!8},
  io/.style={b, dashed},
  data/.style={b, fill=green!12},
  aside/.style={b, fill=gray!12}]
\node[io] (cam) {OV7670\\camera};
\node[unit, right=of cam] (pre) {gray,\\$8\times8$ means};
\node[unit, right=of pre] (bank) {two frame\\banks, $80\times60$};
\node[unit, right=of bank] (ac) {auto-\\contrast};
\node[unit, right=of ac] (eng) {inference engine\\(XNOR, popcount)};
\node[unit, right=of eng] (dec) {decision\\(margin, band)};
\node[io, right=of dec] (host) {host};
\node[data, above=5mm of eng] (model) {on-chip weights,\\thresholds};
\node[aside, below=5mm of bank] (pic) {picture path: SDRAM, VGA};
\node[io, below=5mm of dec] (disp) {VGA overlay,\\LCD, LEDs};
\draw[->] (cam) -- (pre);
\draw[->] (pre) -- (bank);
\draw[->] (bank) -- (ac);
\draw[->] (ac) -- (eng);
\draw[->] (eng) -- (dec);
\draw[->] (dec) -- (host);
\draw[->] (dec) -- (disp);
\draw[->, dashed] (model) -- (eng);
\draw[->] (cam.south) |- (pic.west);
\end{tikzpicture}
\caption{The detection path of the inference HDPU (blue) listens to the camera pins in parallel with the picture path (grey) and needs no processor. The network's weights and thresholds (green) sit in on-chip memory beside the engine. The host, if present, receives a decision of \infoBytes{} bytes per frame.}
\label{fig:arch}
\end{figure*}
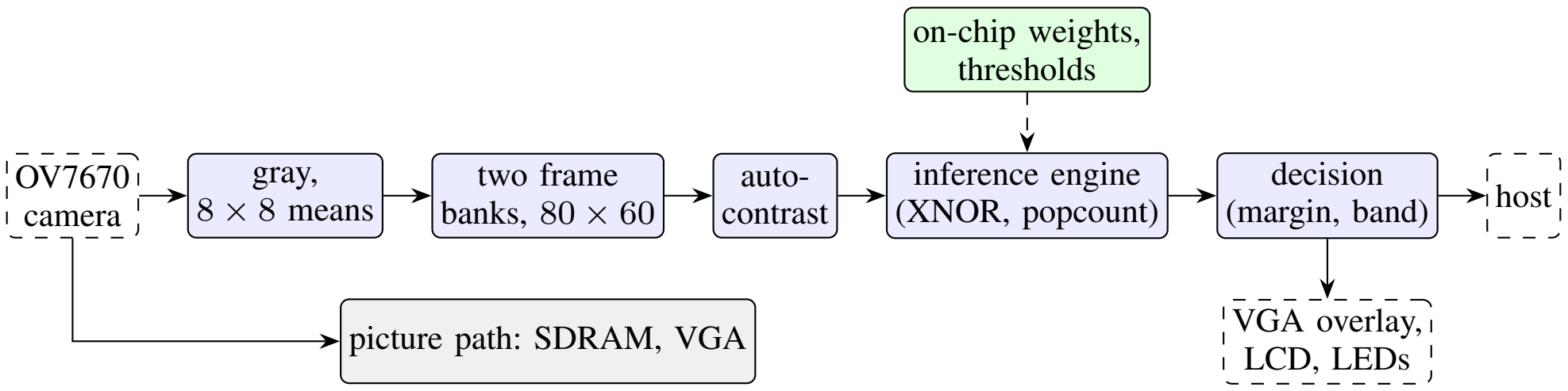

Fig.~\ref{fig:arch} shows the detection path. The OV7670 delivers $640\times480$ RGB565 frames~\cite{ov7670} from a 21.33~MHz clock; with its default VGA timing that is \camFpsModel{} frames per second, or \frameMsModel~ms per frame (model). A pre-processing block converts every pixel to gray as $(77R+150G+29B)\gg8$, exactly as in training, sums $8\times8$ blocks, shifts by six, and writes the resulting $80\times60$ image into one of two alternating frame banks. The sums accumulate as the pixels arrive, so no full-resolution frame is ever stored for detection, and a frame is used only if all of its 480 lines are complete.

\begin{figure}[!htb]
\centering
\includegraphics[width=\columnwidth]{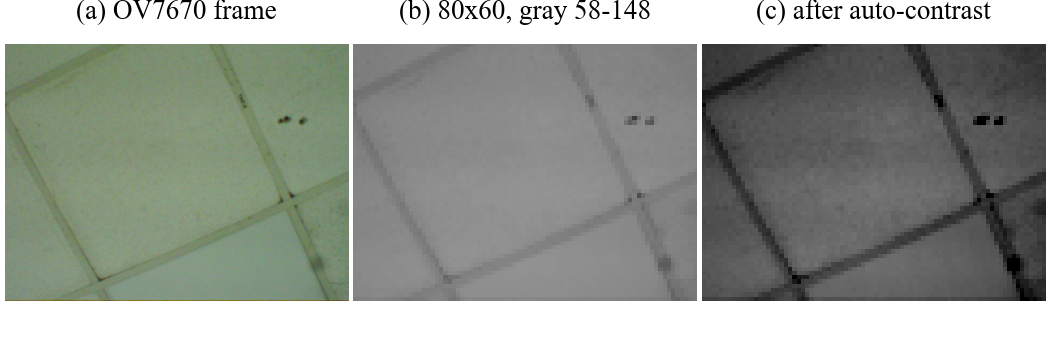}
\caption{A frame captured from the board's OV7670 (a), its $80\times60$ network input, which spans only a narrow range of gray levels (b), and the same input after the hardware auto-contrast stage (c).}
\label{fig:frame}
\end{figure}

The sensor's pictures are flat (Fig.~\ref{fig:frame}). The network was trained on photographs that use the full range of gray levels, so an auto-contrast stage stretches every frame before the network sees it. A 256-bin histogram gives the 25th darkest and the 25th brightest pixel, a 0.5\,\% trim so that a lamp or a dark corner does not defeat the stretch, and the gain is capped at 2.7 so that flat surfaces are not turned into stripe-like texture.

\subsection{The inference engine}
One engine evaluates all five layers in turn (Table~\ref{tab:layers}). The first layer applies binary $3\times3$ weights to the 8-bit pixels, offset by 128. The next three apply binary $3\times3$ weights to binary feature maps by XNOR and population count over all input channels at once, and each is followed by a threshold and a $2\times2$ max-pool, which on binary values is the OR of four bits. The output neuron sums 960 binary products into a score that is compared with a threshold of \vThreeThr. All thresholds come from folding batch normalisation in Python, with negative gains folded into the weights so that every comparison is $\ge$. In all, a decision takes \binaryMacsM{} million binary multiply-accumulates, from \weightBits{} one-bit weights (\weightKB~KB).

The engine walks the loop nest $(p_y, p_x, s, c, k_y, k_x)$ over pooled output positions, the four positions of each pooling window, output channels and kernel taps. It processes one kernel tap per clock in a three-stage pipeline: addresses; a partial sum, which is either a signed pixel or the population count of a 64-bit XNOR; and accumulation, threshold, pooling and write-back. Feature maps are stored as one 64-bit word per pixel, one bit per channel; the weights occupy 2{,}048 words of 64 bits and the thresholds 256 words of 16 bits. A decision therefore takes
\begin{equation}
C = 2 + 4L + \sum_{l=1}^{L} p_{w,l}\,p_{h,l}\,s_l\,c_l\,k_{w,l}\,k_{h,l}
\label{eq:cycles}
\end{equation}
clock cycles, for $L$ layers with $p_{w,l}\times p_{h,l}$ output positions, $s_l=4$ for a pooled layer and 1 otherwise, $c_l$ output channels and a $k_{w,l}\times k_{h,l}$ kernel. The $2+4L$ cycles start the engine, drain the pipeline after each layer and finish. For the deployed network $C=\cyclesInf$, or \inferMs~ms at \bcnnMhz~MHz, and the count depends on the layer table alone, not on the image. Simulation confirms it.

\begin{table}[!htb]
\caption{Layers of the deployed network and the clock cycles each takes in the engine (Eq.~\eqref{eq:cycles})}
\label{tab:layers}
\centering\footnotesize
\resizebox{\columnwidth}{!}{
\begin{tabular}{@{}llrrr@{}}
\toprule
Layer & Operation & Output & Clocks & Share \\
\midrule
L0 & conv $3\times3$, 8-bit input & $40\times30\times32$ & 1{,}382{,}400 & 60.9\,\% \\
L1 & conv $3\times3$, XNOR & $20\times15\times64$ & 691{,}200 & 30.5\,\% \\
L2 & conv $3\times3$, XNOR & $10\times7\times64$ & 161{,}280 & 7.1\,\% \\
L3 & conv $3\times3$, XNOR & $5\times3\times64$ & 34{,}560 & 1.5\,\% \\
L4 & fully connected & score & 15 & 0.0\,\% \\
\midrule
\multicolumn{3}{@{}l}{Total, with 22 control clocks} & 2{,}269{,}477 & \\
\bottomrule
\end{tabular}
}
\end{table}

\subsection{The decision}
The margin $m$ is the score minus the threshold of \vThreeThr. It is averaged over the last \avgFrames{} frames, and the averaged margin gives the unit's state: \emph{crossing} if $m\ge\bandHi$, \emph{no crossing} if $m<-\bandLoAbs$, and \emph{not sure} in between. If no result arrives for one second the state becomes \emph{waiting}. The band was chosen on validation photographs only. It is what lets the unit say ``not sure'' instead of guessing, and Section~\ref{sec:acc} shows how much that is worth.

\section{Verification and Methodology}\label{sec:verif}
\textbf{Bit-exact chain.} A Python integer model computes exactly what the hardware computes, including the pixel offset, the padding values, the thresholds and the pooling. Every level of the design is compared against it (Table~\ref{tab:verif}). In simulation the engine matches the model on all \boardAllN{} test images, in the score, in the decision and in every word of every intermediate feature map. On the board, each test image is sent over the serial link into a dedicated frame bank and classified by the FPGA, and its score and decision are compared with the model's. The self-test also checks 20 golden vectors.

\begin{table}[!htb]
\caption{Verification chain: every level compared with the Python integer model}
\label{tab:verif}
\centering\footnotesize
\resizebox{\columnwidth}{!}{
\begin{tabular}{@{}llr@{}}
\toprule
Level & Compared with the model & Result \\
\midrule
Simulation & camera pins to the $80\times60$ image & 3/3 frames \\
Simulation & engine: score, decision, every feature-map word & 639/639 images \\
Simulation & serial link, contrast, engine, decision (real OV7670 frame) & 20/20 checks \\
Board & self-test, golden vectors over the link & 8/8 checks \\
Board & T1, score and decision & 524/524 images \\
Board & T2, score and decision & 115/115 images \\
\bottomrule
\end{tabular}
}
\end{table}

\textbf{Data.} The data set contains \dataImages{} images, \dataCrossings{} of them with a crossing. From \trainSources{} sources, \splitTrain{} images are used for training, \splitVal{} for validation and \splitTest{} for testing. T1 is the set of \boardTOneN{} dashcam frames from CDSet~\cite{cdset} in the test split, and T2 contains \boardTTwoN{} street photographs by photographers outside the training sources. The network was trained with knowledge distillation from a full-precision ResNet-18~\cite{hinton2015,he2016}.

\textbf{Host.} The host is the Intel Core i5-12500H laptop processor of~\cite{papertwo,intelark}. It ran the network as the PyTorch model used in training, with batch size one under native Windows, after 20 warm-up runs, taking the median of 200 runs, with one thread and with 16 threads. A full-precision ResNet-18 ran under the same conditions. These runs were not controlled as in~\cite{papertwo} (power plan, pinning).

\textbf{Power.} The only power figures are Vivado's vectorless estimates of the routed design, which the tool itself rates as low confidence. No power was measured at the supply, so we compare no energies across platforms.

\textbf{Labels.} Numbers are measured on the board, from simulation, from post-route reports (implementation), from the model, derived from these, or estimated. The text says which whenever it is not obvious.

\section{Results}\label{sec:results}

\subsection{Headline results}
Table~\ref{tab:summary} collects the main results of the unit. The sections that follow give the evidence.

\begin{table}[!htb]
\caption{Headline results of the inference HDPU}
\label{tab:summary}
\centering\footnotesize
\resizebox{\columnwidth}{!}{%
\begin{tabular}{@{}lrl@{}}
\toprule
Result & Value & Source \\
\midrule
Decision time, every frame, any image & \inferMs~ms (\cyclesInf{} clocks) & simulation \\
Frames skipped on the live camera & 0 & hardware \\
Score and decision equal to the model & \boardAllExact{} of \boardAllN{} images & hardware \\
Accuracy, T1 / T2 & \boardTOneAcc{} / \boardTTwoAcc\,\% & hardware \\
Correct when answering, T1 / T2 & \confTOneAcc{} / \confTTwoAcc\,\% & hardware \\
Camera-like frames, T1: without $\rightarrow$ with contrast stage & \camlikeTOneRaw\,\% $\rightarrow$ \camlikeTOneSt\,\% & model \\
Logic, detection path / whole system & \detectLut{} / \sysLut{} LUTs & implementation \\
Host time spared per frame (one core) & \cpuBcnnOneMs~ms, \hostSharePct\,\% of a frame & host \\
Data to the host per frame & \infoBytes{} B, \frameReduction$\times$ less & derived \\
\bottomrule
\end{tabular}}
\end{table}

\subsection{On the board}\label{sec:board}
Table~\ref{tab:board} summarises the board results. The unit evaluated every frame of the live camera: its on-chip counter registered \boardInferRate{} decisions in a one-second window, with no frame skipped. It returned exactly the scores and decisions of its software model on all \boardAllN{} test images, so the hardware reproduces the accuracy measured in software.

\begin{table}[!htb]
\caption{Measured on the EDGE Artix-7 board}
\label{tab:board}
\centering\footnotesize
\begin{tabular}{@{}ll@{}}
\toprule
Check & Result \\
\midrule
Self-test & 8 of 8 checks pass \\
Camera frames committed & 27.5 frames/s, none rejected \\
Decisions on the live camera & 27/s, no frame skipped \\
T1, dashcam frames, through the FPGA & 524/524 equal to the model, 79.39\,\% \\
T2, street photos, through the FPGA & 115/115 equal to the model, 79.13\,\% \\
\bottomrule
\end{tabular}

\end{table}

\subsection{A fixed decision time inside the frame budget}\label{sec:latency}
\begin{figure*}[t]
\centering
\includegraphics[width=\textwidth]{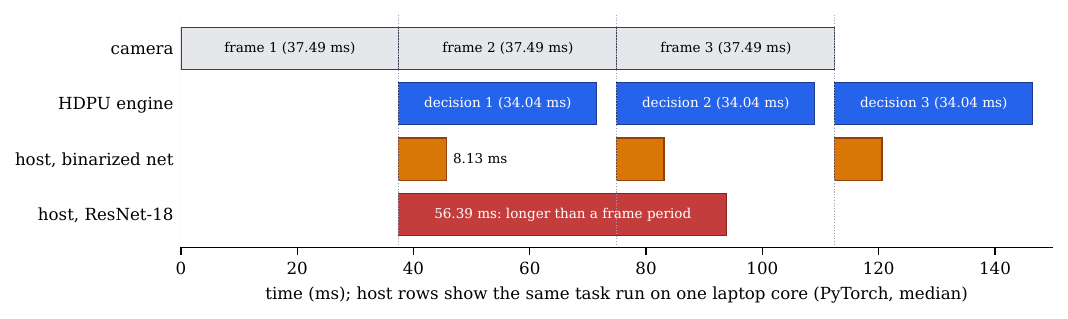}
\caption{Three consecutive frame periods to scale. The HDPU engine (blue) finishes each decision inside the next frame period, never missing a frame. The two host rows show how long the same task takes on one laptop core (median, PyTorch, Section~\ref{sec:verif}), not a measured sequence: the binarized network takes \cpuBcnnOneMs~ms, and the full-precision ResNet-18 takes longer than a frame period.}
\label{fig:timeline}
\end{figure*}

Every decision takes \cyclesInf{} clock cycles, \inferMs~ms, whatever the image (Eq.~\eqref{eq:cycles}; simulation). Nothing in the unit depends on the image content, on other tasks or on history, so the decision time is a property of the design. It is also shorter than the camera's frame period of \frameMsModel~ms, which leaves \slackMs~ms to spare in every frame (Table~\ref{tab:budget}). The engine is busy \busyPct\,\% of the time, never falls behind the camera, and starts its next decision as soon as the next frame is committed (Fig.~\ref{fig:timeline}). Counting from the end of a frame, including the contrast stage, the decision follows after about \commitToDecisionMs~ms (derived).

\begin{table}[!htb]
\caption{Frame budget of the inference HDPU}
\label{tab:budget}
\centering\footnotesize
\begin{tabular}{@{}lrl@{}}
\toprule
Quantity & Value & Source \\
\midrule
Camera frame period & 37.49~ms & model \\
Decision time & 34.04~ms & simulation \\
Margin per frame & 3.44~ms & derived \\
Frame end to decision & about 34.2~ms & derived \\
Engine busy & 90.8\,\% of the frame & derived \\
Highest decision rate & 29.4/s (camera: 26.68/s) & derived \\
Frames skipped on the board & 0 & hardware \\
\bottomrule
\end{tabular}

\end{table}

\subsection{Accuracy and answers}\label{sec:acc}
Table~\ref{tab:decisions} gives the decisions of the unit on both test sets, identical to the model's. On the dashcam frames it is conservative (precision \precisionTOne\,\%, recall \recallTOne\,\%); on the street photographs it is balanced (precision \precisionTTwo\,\%, recall \recallTTwo\,\%).

\begin{table}[!htb]
\caption{Decisions of the unit on the held-out test sets}
\label{tab:decisions}
\centering\footnotesize
\begin{tabular}{@{}lrr@{}}
\toprule
 & T1 (dashcam) & T2 (street) \\
\midrule
Images (with a crossing) & 524 (292) & 115 (67) \\
True / false detections & 197 / 13 & 56 / 13 \\
Missed / true rejections & 95 / 219 & 11 / 35 \\
Precision / recall (\%) & 93.8 / 67.5 & 81.2 / 83.6 \\
F1 / accuracy (\%) & 78.5 / 79.4 & 82.4 / 79.1 \\
Answered outside the band (\%) & 50.8 & 40.0 \\
Correct among answered (\%) & 90.2 & 95.7 \\
\bottomrule
\end{tabular}

\end{table}

\begin{figure}[!htb]
\centering
\includegraphics[width=\columnwidth]{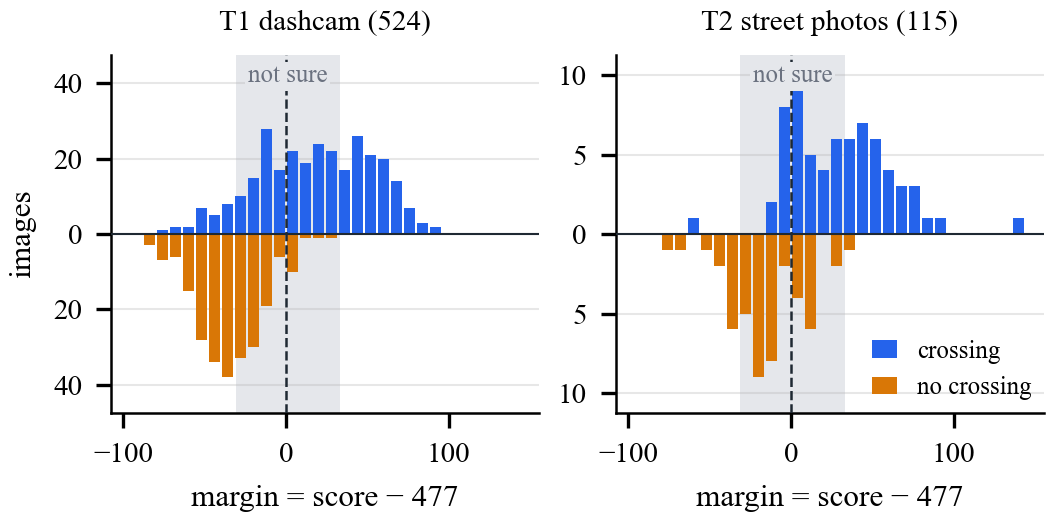}
\caption{Margin (score minus the output threshold of \vThreeThr) of every test image, from the bit-exact model; the board returned the same scores for all \boardAllN{} images. Crossings above and non-crossings below the axis; grey: the not-sure band.}
\label{fig:margins}
\end{figure}

\begin{figure}[!htb]
\centering
\includegraphics[width=\columnwidth]{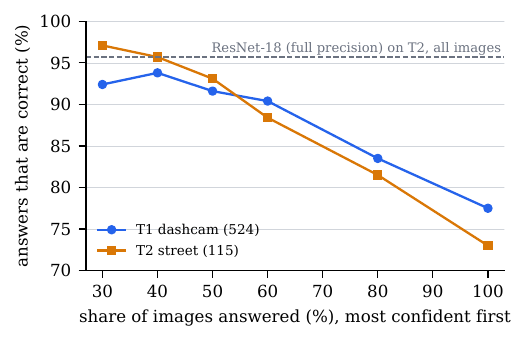}
\caption{Share of correct answers against the share of images answered, most confident first, through the camera path (model). The dashed line is the full-precision ResNet-18 on all of T2.}
\label{fig:coverage}
\end{figure}

The not-sure band is what turns an average classifier into a reliable one. Fig.~\ref{fig:margins} shows why: most of the overlap between the two classes falls inside the band, so withholding an answer there removes most of the errors. When the unit does answer, which it does for \confTOneCov\,\% of the images of T1 and \confTTwoCov\,\% of T2, \confTOneAcc\,\% and \confTTwoAcc\,\% of its answers are correct. Ranking every image by confidence extends this to any coverage (Fig.~\ref{fig:coverage}, Table~\ref{tab:selective}): on street photographs the most confident 30\,\% of the images are \selPredTTwoThirty\,\% correct, and on dashcam frames the most confident 40\,\% are \selPredTOneForty\,\% correct. A full-precision ResNet-18 on the same $80\times60$ input reaches \resnetTTwo\,\% on T2, so binarization, and not the input, limits accuracy.

\begin{table}[!htb]
\caption{Accuracy of the most confident answers through the camera path (model)}
\label{tab:selective}
\centering\footnotesize
\resizebox{\columnwidth}{!}{
\begin{tabular}{@{}lrrrrrr@{}}
\toprule
Share answered (\%) & 30 & 40 & 50 & 60 & 80 & 100 \\
\midrule
T1 (dashcam), correct (\%) & 92.4 & 93.8 & 91.6 & 90.4 & 83.5 & 77.5 \\
T2 (street), correct (\%) & 97.1 & 95.7 & 93.1 & 88.4 & 81.5 & 73.0 \\
\bottomrule
\end{tabular}
}
\end{table}

\subsection{The real camera and the contrast stage}\label{sec:camera}
\begin{figure}[!htb]
\centering
\includegraphics[width=\columnwidth]{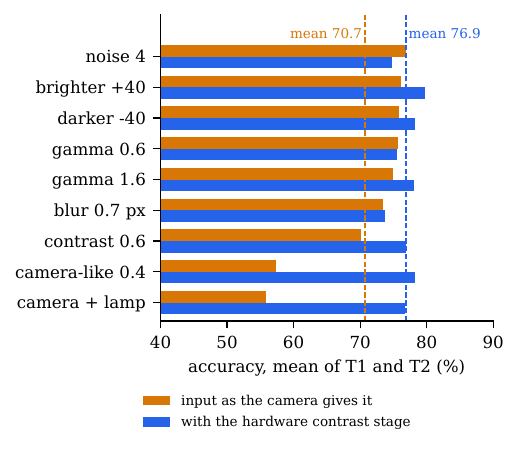}
\caption{Accuracy under nine camera effects, mean of T1 and T2 (model), with the network input as the camera gives it and with the hardware contrast stage.}
\label{fig:robust}
\end{figure}

The sensor matters as much as the network. A frame captured from the board's camera spans only a narrow range of gray levels (Fig.~\ref{fig:frame}), a range the network never saw in training. On dashcam frames degraded to the camera's contrast, accuracy falls to \camlikeTOneRaw\,\% (T1) and \camlikeTTwoRaw\,\% (T2), and the hardware contrast stage restores \camlikeTOneSt\,\% and \camlikeTTwoSt\,\%. It does so across the board (Fig.~\ref{fig:robust}): averaged over nine camera effects (contrast, brightness, gamma, blur, noise and a lamp in view), accuracy rises from \camAvgRaw\,\% to \camAvgStretch\,\%, and the trimmed histogram keeps the gain when a lamp is in view, where a plain min--max stretch fails. The stage costs a small loss on images that already use the full range, which is why photographs sent from a PC bypass it, and it takes an estimated 0.15~ms.

\subsection{Cost}\label{sec:cost}
\begin{figure*}[t]
\centering
\includegraphics[width=\textwidth]{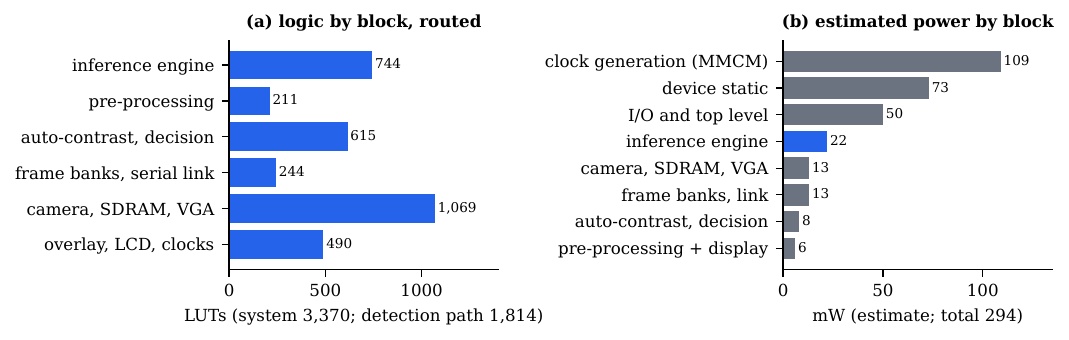}
\caption{(a) LUTs by block of the routed design. (b) Vivado's vectorless power estimate by block; the inference engine (blue) is a small part of the total.}
\label{fig:cost}
\end{figure*}

Table~\ref{tab:res} and Fig.~\ref{fig:cost} give the cost. The whole detection path, from pre-processing to the serial link, uses \detectLut{} LUTs, and the inference engine itself \coreLut{} LUTs and \coreBram{} block RAMs, with no DSP slice. The complete system, including the camera interface, the picture path and the displays, uses \sysLut{} LUTs (\sysLutPct\,\% of the device), \sysBram{} of its \devBram{} block RAMs and \sysDsp{} DSP slice. All clocks meet timing after routing, the engine's clock with \bcnnWns~ns of setup slack at \bcnnMhz~MHz.

Vivado estimates the complete design at \powTotalW~W, \powDynW~W dynamic and \powStaticW~W static. The engine accounts for only \corePowMw~mW of it, or \coreEnergyMj~mJ per decision (derived from the estimate); clock generation (\mmcmPowMw~mW) is the largest single part. Since no power was measured at the supply, we compare no energies across platforms.

\begin{table}[!htb]
\caption{Resources on the XC7A35T-1 (Vivado 2024.2, routed). $^{*}$From the hierarchical report; LUT sharing makes the sum of the rows slightly larger than the total.}
\label{tab:res}
\centering\footnotesize
\resizebox{\columnwidth}{!}{
\begin{tabular}{@{}lrrrr@{}}
\toprule
Block & LUT & FF & BRAM36 & DSP \\
\midrule
Inference engine (BCNN core) & 744 & 321 & 10 & 0 \\
Pre-processing (gray, $8\times8$ means) & 211 & 257 & 0 & 0 \\
Auto-contrast, loader, decision & 615 & 593 & 0 & 1 \\
Frame banks, serial link & 244 & 132 & 7 & 0 \\
Camera, SDRAM, VGA picture path & 1{,}069 & 769 & 1.5 & 0 \\
Overlay, LCD, buzzer, clocks & 490 & 658 & 0.5 & 0 \\
\midrule
Complete system$^{*}$ & 3{,}370 & 2{,}730 & 19 & 1 \\
\bottomrule
\end{tabular}
}
\end{table}

\subsection{What the host is spared}\label{sec:host}
The decision of the unit is work that the host never does. Running the same binarized network as the PyTorch model used in training takes \cpuBcnnOneMs~ms of one host core in every frame period, \hostSharePct\,\% of it, and \hostSixteenSharePct\,\% with 16 threads (Table~\ref{tab:host}). A full-precision ResNet-18, which reaches \resnetTTwo\,\% on T2, needs more than a frame period: \cpuResnetOneMs~ms on one thread and \cpuResnetSixteenMs~ms on all 16, against \frameMsModel~ms (derived from medians). An inference HDPU removes all of this from the host, and with it the transfer of the frame.

\begin{table}[!htb]
\caption{Host time for the same inference (PyTorch, batch 1, median of 200 runs; uncontrolled), against the inference HDPU}
\label{tab:host}
\centering\footnotesize
\resizebox{\columnwidth}{!}{
\begin{tabular}{@{}lrrr@{}}
\toprule
Host task per frame & 1 thread & 16 threads & Share of frame, 1 thread \\
\midrule
Binarized network & 8.13~ms & 10.47~ms & 21.7\,\% \\
ResNet-18, full precision & 56.39~ms & 42.49~ms & 150\,\% \\
\midrule
Inference HDPU & \multicolumn{2}{r}{no host time} & 0\,\% \\
\bottomrule
\end{tabular}
}
\end{table}

\begin{figure}[!htb]
\centering
\includegraphics[width=\columnwidth]{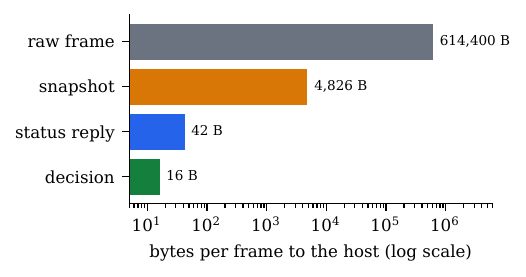}
\caption{Bytes per frame that the host would receive, on a log scale.}
\label{fig:data}
\end{figure}

The saving in data is just as large. To decide from a raw frame, the host must first receive \frameBytes{} bytes. The unit's decision is \infoBytes{} bytes, \frameReduction{} times less, and even the image the network saw together with its decision (\snapshotBytes{} bytes) is \snapshotReduction{} times smaller than the frame (Table~\ref{tab:interface}, Fig.~\ref{fig:data}).

\begin{table}[!htb]
\caption{Messages from the unit to the host, per frame (derived from the serial protocol)}
\label{tab:interface}
\centering\footnotesize
\resizebox{\columnwidth}{!}{
\begin{tabular}{@{}lrr@{}}
\toprule
Message to the host & Bytes & Smaller than a frame by \\
\midrule
Raw RGB565 frame & 614{,}400 & 1$\times$ \\
Snapshot: the $80\times60$ image the network saw, with its decision & 4{,}826 & 127$\times$ \\
Status reply & 42 & 14{,}629$\times$ \\
Decision (state, score, margin, average, counters) & 16 & 38{,}400$\times$ \\
\bottomrule
\end{tabular}
}
\end{table}

\subsection{Positioning}\label{sec:position}
Table~\ref{tab:compare} sets the unit beside published zebra-crossing detectors and a binarized-network accelerator. Each reports on its own data and task, so the accuracies cannot be ranked. What differs is what each needs around it. The published detectors run on a phone or a GPU module with an operating system, and FINN's prototype on a Zynq SoC board, whereas this unit needs only its board and the camera. It has no processor and no operating system, it decides on every frame, and it leaves the host free.

\begin{table*}[t]
\caption{Positioning against published systems, as reported by their authors. Accuracies refer to each system's own data and task. $^{\dagger}$Reported for the classification of pedestrian traffic lights.}
\label{tab:compare}
\centering\footnotesize
\begin{tabular}{@{}lllll@{}}
\toprule
System & Platform & What it needs around it & Rate & Reported result \\
\midrule
ZebraRecognizer~\cite{mascetti2016} & iPhone 5S & phone, operating system & about 25 fps & precision 1.0, recall 0.93 \\
LYTNet~\cite{yu2019} & iPhone 7 & phone, operating system & 20 fps & 94\,\%$^{\dagger}$ \\
CDNet~\cite{zhang2022} & Jetson Nano & GPU module, operating system & 33.1 fps & F1 above 94\,\% \\
FINN CNV~\cite{umuroglu2017} & ZC706 (Zynq-7045) & Zynq SoC board & 21{,}906 images/s & 80.1\,\% (CIFAR-10) \\
\textbf{This work} & XC7A35T and a camera & nothing else: no processor, no OS & \camFpsModel{} fps (every frame) & \vThreeTTwo\,\% (T2); \confTTwoAcc\,\% when answering \\
\bottomrule
\end{tabular}
\end{table*}

\section{Discussion}
\textbf{From pre-processing to decisions.} The units of~\cite{paperone,papertwo} turned the pixel stream into edges, corners and a steering code. This unit turns it into a decision. They all share one organisation: a fixed datapath at the sensor, a time fixed by design, and a compact result for the host. HDPUs do not compete for a bus or a scheduler, so an inference unit can in principle sit beside the Canny, Harris and line units on one camera stream, and the host would then receive edges, corners, a steering code and a decision, instead of frames.

\textbf{What the result shows.} The unit gives a neural decision the same properties as the pre-processing units before it: it decides at a time that does not depend on the image, it needs no processor, and it takes its work off the host. The contrast stage shows that these properties extend to the sensor side of the decision too. A learned decision is only as good as the signal it receives, and the cheap hardware stage in front of the network recovers most of what the real sensor costs.

\textbf{Limitations.} The power figures are Vivado estimates; no power was measured at the supply. The frame period is the model value of the camera's register settings, and the board's counters agree with it to within their one-frame resolution. The decision time is exact by construction and in simulation, but was not timed on the board with counters. The host runs were not controlled, and used PyTorch. T2 has \boardTTwoN{} images, so its 95\,\% confidence interval is about $\pm7$ points, and the live-camera tests are qualitative. The paper makes no claim about inference speed on images already stored in memory. Board power, on-chip latency counters, a controlled host baseline and an inference unit running beside the pre-processing units of~\cite{papertwo} on one camera stream are the next steps.

\section{Conclusion}
An inference HDPU takes a machine-learning decision at the sensor with no processor in the loop. On a low-cost Artix-7 FPGA it decides on every camera frame in a fixed \cyclesInf{} clock cycles, matches its software model on all \boardAllN{} test images on the board, and uses \detectLut{} LUTs from pixels to decision. Its hardware contrast stage lifts accuracy on camera-like frames from \camlikeTOneRaw\,\% to \camlikeTOneSt\,\%, and its not-sure band makes \confTTwoAcc\,\% of its street-photo answers correct. The host runs none of the inference and receives \infoBytes{} bytes per frame instead of a frame. Together with the pre-processing units before it~\cite{paperone,papertwo}, this shows that hardwired distributed processing units at the sensor can take on both the preparation and the decisions of a vision system, and leave the host free for the work that needs it.

\bibliographystyle{IEEEtran}
\bibliography{refs}

\end{document}